\documentclass[]{tPHM2e}

\DeclareGraphicsExtensions{.ps,.eps,.EPSF} 

\begin{document}

\newcommand{\spd}{$sp$--$d$ }
\newcommand{\ef}{E_{\rm F}}
\newcommand{\du}{{\rm d}}
\newcommand{\e}{{\rm e}}
\newcommand{\Ang}{{\rm \AA}}

\newcommand{\be}{\begin{equation}}
\newcommand{\ee}{\end{equation}}
\newcommand{\ben}{\begin{eqnarray}}
\newcommand{\een}{\end{eqnarray}}
\newcommand{\beq}{\begin{equation}}
\newcommand{\eeq}{\end{equation}}
\newcommand{\B}{\mathrm{B}}
\newcommand{\NB}{\mathrm{NB}}
\newcommand{\RRe}{\mathrm{Re}}
\newcommand{\IIm}{\mathrm{Im}}

\doi{10.1080/14786435.20xx.xxxxxx}
\issn{1478-6443}
\issnp{1478-6435}

\markboth{G. Trambly de Laissardi\`ere, C. Oguey, D. Mayou}
{Quantum transport in octagonal tilling approximants}


\title{
Breakdown of the semi-classical conduction theory 
in approximants of the octagonal tiling
}

\author{G. Trambly de Laissardi\`ere,$^{\dagger}$\thanks{$^\ast$Corresponding author.
Email: guy.trambly@u-cergy.fr}$^\ast$
C. Oguey$^{\dagger}$
and D. Mayou$^{\ddagger}$\\\vspace{6pt}
$^{\dagger}$ Laboratoire de Physique Th\'eorique et Mod\'elisation,
CNRS UMR 8089, Universit\'e de Cergy-Pontoise,
2 av. A. Chauvin, 95302 Cergy-Pontoise, France
\\
$^{\ddagger}$ Institut N\'eel, CNRS, Universit\'e Joseph Fourier, B\^at D,
25 av. des Martyrs,\\
B.P. 166, 38042 Grenoble Cedex 9, France\\
\received{August 2010}}

\maketitle

\begin{abstract}
We present numerical calculations of quantum transport in perfect octagonal approximants.
These calculations include a Boltzmann (intra-band) contribution
and a non-Boltzmann  (inter-band) contribution.
When the unit cell size of the approximant increases, the magnitude of Boltzmann terms decreases,
whereas the magnitude of non-Boltzmann terms increases.
It shows that, in large approximants, the
non-Boltzmann contributions should dominate the transport properties of electrons.
This confirms the break-down of the Bloch-Boltzmann theory to understand the
transport properties in approximants with very large unit cells, and then in quasicrystals,
as found in actual Al-based approximants.

\end{abstract}

\section{Introduction}

\vskip .5cm
Since the discovery of  Shechtman et al. \cite{Shechtman84}
many
experimental investigations have indicated that the conduction
properties of several stable quasicrystals (AlCuFe, AlPdMn, AlPdRe...) are at the opposite of those of
good metals
\cite{Poon92,Berger94}.
It appears also that the medium range order, over one or a few nanometres, is the real length scale
that determines conductivity. 
In particular, the role of transition elements enhancing localisation has been often studied 
\cite{Pasturel86,Fujiwara89,Fujiwara93,Krajci95,Fujiwara96,cluster97,cluster97b,revue_HR}.
There is now strong evidence that these non standard properties
result from a new type of break-down of the semi-classical Bloch-Boltzmann theory of conduction
\cite{PRL06,JubileeQC}.
On the other hand, the specific role of long range quasiperiodic order on transport properties is
still an open question in spite of a large number of studies (Refs.  
\cite{Sire90,Passaro92,Jagannathan94,Zhong94,Roche97,Zijlstra02,Zijlstra04,Jagannathan07,Triozon02,Yamamoto95,Fujiwara96} 
and Refs. therein).
In this paper, we study ``how electrons propagate''
in approximants of the octagonal tiling. This tiling is one of the well-known  quasiperiodic tilings
that have been used to understand the influence of quasiperiodicity on electronic transport
\cite{Sire90,Passaro92,Jagannathan94,Zhong94,Zijlstra02,Zijlstra04,Jagannathan07}.
The main objective is to show that non standard conduction properties result from
purely quantum effects due to quasiperiodicity that cannot be interpreted
through the semi-classical theory of transport.

\section{The octagonal tiling and its approximants}
The octagonal, or Ammann-Beenker, tiling \cite{Ammann92,Socolar89} is a quasiperiodic 
tiling analogous to the notorious Penrose tiling.
It has eightfold C$_{\rm 8v}$ point symmetry, inflation symmetry with ratio $\lambda=1+\sqrt2$,
quasiperiodic translation symmetry, and Bragg peaks indexed by $\mathbb{Z}^4$.
Those symmetries were observed in quasicrystalline phases of CrNiSi and AlMnSi alloys \cite{Wang88}.

A sequence of periodic approximants $X_0, X_1,\ldots,X_k,\ldots$ can be generated
by the inflation mapping $M$, operating as an invertible integer matrix on  $\mathbb{Z}^4$
\cite{DuMoOg89}. In the physical Euclidean plane $E$, and in the simplest version
where the atoms occupy the tiling vertexes, all the (atomic) sites have integer
coordinates in the basis $(e_1, e_2, e_3, e_4)$, where  $(e_1, e_2)$ forms an orthogonal
basis of $E$ and $(e_3, e_4)$ is another orthogonal basis related to the former by 45$^\circ$
rotation. All the tiles are congruent to either a square or a rhombus with 45$^\circ$ acute angle.

For $k=0,1,2,\ldots$, the $k^\mathrm{th}$ approximant is periodic, invariant under
the translation lattice $\mathcal{L}_k$ generated by $\lambda^k e_1$ and $\lambda^k e_2$
(the square lattice  $[e_1, e_2]$ dilated by a factor $L_k=\lambda^k$). In terms of the Pell,
or ``octonacci'', number sequence $P = \{P_k\} = \{0,1,2,5,12,\ldots\}$, defined by
$P_0=0,\ P_1=1,\ P_{k+1}=2P_k+P_{k-1}$, the unit cell contains $N_k=P_{2k+1}+P_{2k}$ vertexes.
Those vertexes are given by the following construction: $\lambda^k(W_k \cap [e_3, e_4])$,
\textit{i.e.} select and expand sites of the square lattice $[e_3, e_4]$, eventually bring
them back into the square fundamental domain by $\mathcal{L}_k$ translations. The selection window $W_k$
is a semi regular octagon obtained as a projection of the 4D unit cube. Up to a sign
$(-1)^k$, $W_k$ is the span (all combinations with coordinates between 0 and 1) of the four
vectors
$$\begin{array}{llll}
P_k (e_4 - e_3),& -P_k (e_3 + e_4),&
(P_k+P_{k-1}) e_3,& (P_k+P_{k-1}) e_4.
\end{array}$$

Ref. \cite{DuMoOg89} provides a slightly more efficient numerical method, avoiding useless scans,
by partitioning the octagon into six parallelogram sub-domains; then each selection reduces to
specifying ranges for the integer coordinates of a suitable lattice, followed by a global affine
transformation.

\begin{figure}
  \centering
  \includegraphics[height=4cm]{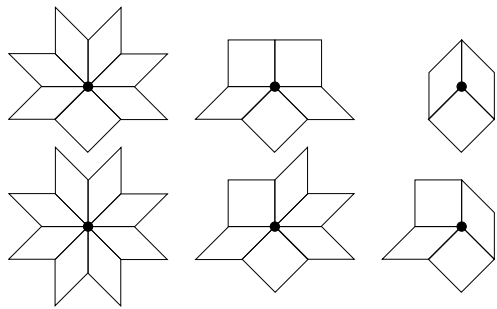}  
  \caption{Vertex configurations of the octagonal tiling and of its approximants $k\geq1$.}
  \label{vertexconf}
\end{figure}

The first approximant $X_0$ is a simple square tiling. In all subsequent approximants
$k\geq 1$, the configurations around vertexes are the same as in the octagonal quasiperiodic
tiling, displayed in Fig. \ref{vertexconf}. Only the approximants $X_1$
contain incomplete subsets of these six local patterns, depending on the phase.

In units of edge length $a=||e_j||$, 
the shortest distances between vertexes are
 $1/\sqrt 2, 1, \sqrt 2$, corresponding to the short rhombus diagonal,
the tile edge and the square diagonal respectively.

\section{Electronic structure}

\begin{figure}[]
\begin{center}
\includegraphics[width=8cm]{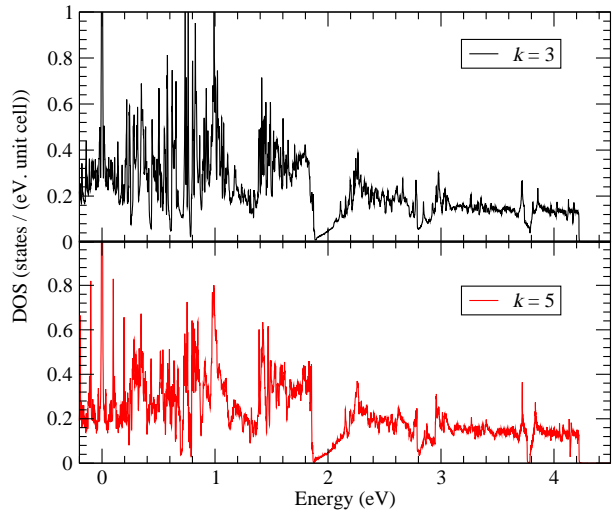}
\caption{\label{Fig_DOS}
(colour online)
Total density of states in perfect octagonal approximants $k=3$ and $k=5$.
DOS is symmetric w.r.t. $E=0$ \cite{Zijlstra04}.
}
\end{center}
\end{figure}

We study a pure hopping Hamiltonian
\ben
 H ~=~  \gamma ~ \sum_{\langle i,j \rangle} | i  \rangle \langle j |
\label{Eq_pureHopHamilt}
\een
where $i$ indexes s orbitals located on all vertexes and $\gamma = 1\,$eV 
is the strength of the hopping between orbitals.
$\langle i,j \rangle$ are the nearest-neighbours at tile edge distance $a$
(figure \ref{vertexconf}). 
The properties of this model depend only on topology of the
tiling.
The total density of states (total DOS), $n(E)$, is computed by a recursion method  
\cite{Triozon02} in a
large super-cell containing more than about 10 million orbitals. 
DOS is shown in figure \ref{Fig_DOS}.
As already shown by E. S. Zijlstra \cite{Zijlstra04}, DOS is spiky
at the centre of the band ($|E|<2$) and smooth
near the band edges ($|E|>2$).

\begin{figure}[]
\begin{center}
\includegraphics[width=8cm]{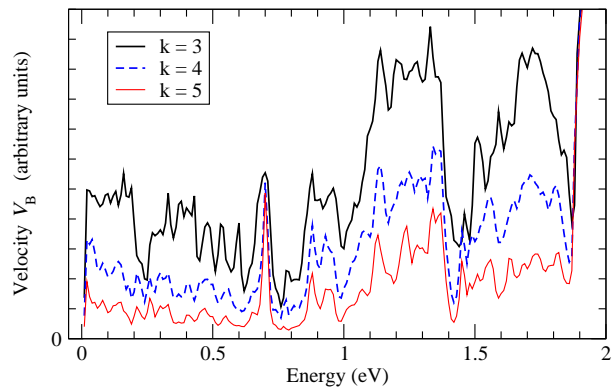}
\caption{\label{Fig_VB}
(colour online)
Average Boltzmann velocity $V_{\B}$ in perfect octagonal approximants $k=3$, $k=4$ and $k=5$.
}
\end{center}
\end{figure}

\begin{figure}[]
\begin{center}
\includegraphics[width=8cm]{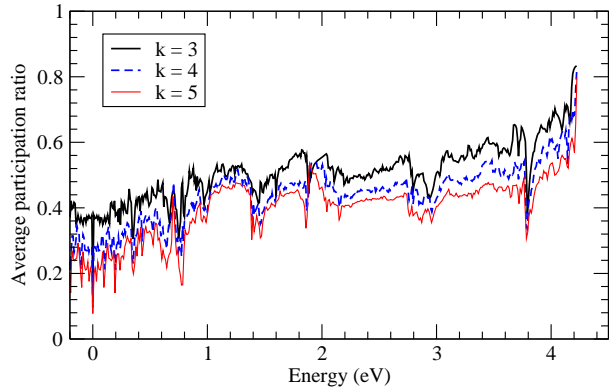}
\caption{\label{Fig_PartRatio}
(colour online)
Average participation ratio $\tilde{p}(E)$ in perfect octagonal approximants $k=3$, $k=4$ and $k=5$.
}
\end{center}
\end{figure}

The Boltzmann velocity (intra-band velocity) 
$V_{\B} \propto \frac{\partial E_n(q)}{\partial q}$ ($q$: wave vector) is shown
figure \ref{Fig_VB}.
When the size of the approximant increases, $V_\B$ decreases as expected from band scaling
analysis \cite{Sire90}.
In order to quantify this localisation phenomenon, we
compute  the participation ratio of each  eigenstate $| \psi \rangle$
defined by
\ben
p(\psi) = \Big( N \sum_{i=1}^N |\langle i | \psi \rangle|^{4} \Big)^{-1}\,,
\een
where 
$i$ indexes orbitals in a  $\mathcal{L}_k$ unit cell
and $N (\equiv N_k)$ is the number of atoms in this unit cell. 
For completely delocalised eigenstates $p$ is equal to $1$.
On the other hand, states localised on one site have a small
$p$ value: $p=1/N$.
The average participation ratio $\tilde{p}(E)$ of all states with energy close to
$E$ are presented on
figure \ref{Fig_PartRatio}.
This figure shows clearly a stronger localisation of electronic states for larger
approximants (larger $k$ values).

\section{Quantum diffusion}

In the framework of Kubo-Greenwood approach for calculation of the conductivity,
the central quantities are the velocity correlation function of
states of energy $E$ at time $t$,
\begin{eqnarray}
C(E,t) = \Big\langle V_x(t)V_x(0) + V_x(0)V_x(t) \Big\rangle_E
= 2\,{\rm Re}\, \Big\langle V_x(t)V_x(0) \Big\rangle_E \, ,
\label{EqAutocorVit}
\end{eqnarray}
and the average
square spreading of states of energy $E$ at time $t$
along the $x$ direction,
\begin{eqnarray}
  \Delta X^{2}(E,t) =\Big< \Big(X(t)-X(0) \Big)^{2}\Big>_{E} .
 \label{Def_2}
 \end{eqnarray}
In (\ref{EqAutocorVit}),  Re$\,A$ is the real part of $A$ and
$V_x(t)$ is the Heisenberg representation of the velocity operator $V_x$
along $x$ direction at time $t$:
\begin{eqnarray}
V_x = \frac{1}{i \hbar}~ \Big[ X , H \Big].
\end{eqnarray}
$C(E,t)$ is related to quantum
diffusion by:
\begin{eqnarray}
\frac{\rm d}{{\rm d} t} \Big(\Delta X^2(E,t) \Big) = \int_0^{t}C(E,t'){\rm d} t'.
\label{EqX2}
\end{eqnarray}
At zero temperature, the static conductivity is given by the Einstein formula,
\ben
\sigma = e^2 n(E_{\rm F}) D(E_{\rm F})\, ,
\een
where $E_{\rm F}$ is the Fermi energy 
and $D$, the diffusivity. In a perfect tiling at temperature $T=0$\,K,
\ben
 D(E) = \frac{\Delta X^2(E,t)}{t} \quad (\mathrm{as}\  t \rightarrow +\infty) .
\label{EqDiffusion}
\een
Once the band structure is calculated from the tight-binding Hamiltonian
(\ref{Eq_pureHopHamilt}) the
velocity correlation function can be computed exactly in the basis
of Bloch states \cite{PRL06,Mayou08RevueTransp}.
Relation (\ref{EqX2}) shows that an anomalous
behaviour of $C(E,t)$ also implies an anomalous behaviour of the
quantum diffusion.
In crystalline approximants, at the long time, when the spreading 
of states, $\Delta X = \sqrt{\Delta X^2}$, reaches the size  $L_k$ of a unit cell,
the propagation should become 
ballistic. This means that the time averages at long times are
$C(E,t) \simeq 2 V_{\B}^2$
and
$\Delta X^2(E,t) \simeq V_{\B}^2 t^2$, 
with $V_{\B}$ the Boltzmann velocity along the $x$ direction.
But for smaller time values, when $\Delta X < L_k$, the spreading  of states depends 
only on the tiling structure within a unit cell, which is closely related 
to quasiperiodicity. 
Therefore, an analysis of electron propagation
in approximants  on time range before the ballistic regime gives a good indication on
how electrons propagate in quasiperiodic structure.

\begin{figure}[]
\begin{center}
\includegraphics[width=8cm]{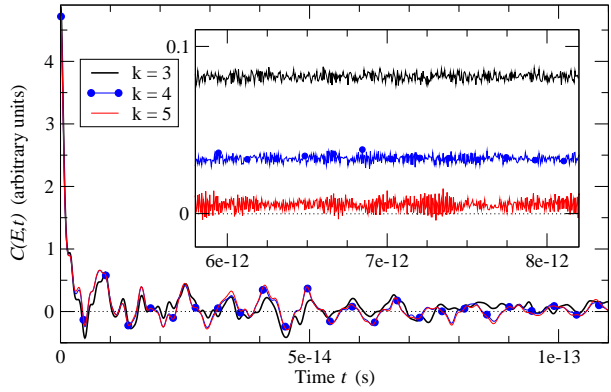}
\caption{\label{Fig_VX2}
(colour online)
Velocity correlation function $C(E,t)$ versus time $t$
in perfect octagonal approximants $k=3$, $k=4$ and $k=5$.
For $E=0.5$\,eV.
}
\end{center}
\end{figure}
As shown on figure \ref{Fig_VX2},  the velocity correlation function in approximants
can take negative values at some times, indicating the presence of back-scattering effects in
electron propagation.
The decay of the Boltzmann term when $k$ increases
suggests that, for larger approximants,
back-scattering ($C(E,t)<0)$ should be obtained even at some large time values.
In that case, conductivity of the tiling will be strongly affected
by this back-scattering phenomenon (equation \ref{EqDiffusion}).

\begin{figure}[]
\begin{center}
\includegraphics[width=8cm]{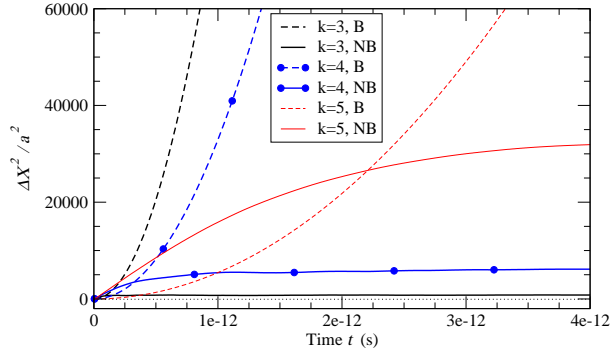}
\caption{\label{Fig_DX2_E.5}
(colour online)
Boltzmann ($\Delta X^2_{\B}$) and Non-Boltzmann ($\Delta X^2_{\NB}$) square spreading versus time $t$
at $E=0.5$\,eV
in perfect octagonal approximants $k=3$, $k=4$ and $k=5$.
$a$ is the length of edges of the tiling.
}
\end{center}
\end{figure}
\begin{figure}[]
\begin{center}
\includegraphics[width=8cm]{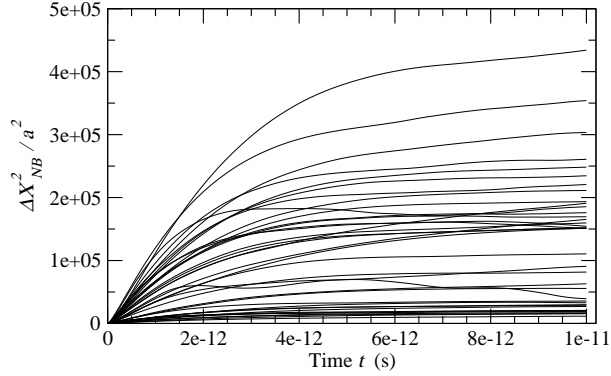}
\caption{\label{Fig_DX2_NB}
Non-Boltzmann square spreading $\Delta X^2_{\NB}$ versus time $t$ at different energies $E$
in perfect octagonal approximant $k=5$.
$a$ is the tile edge length.
}
\end{center}
\end{figure}

The average square spreading is the sum of two terms (figure \ref{Fig_DX2_E.5}):
\begin{equation}
\Delta X^2(E,t) = V_{\B}^2 t^2 + \Delta X_{\NB}^2(E,t).
\end{equation}
The first term is the ballistic (intra-band)  contribution
at the energy $E$.
The semi-classical model of Bloch-Boltzmann theory
is equivalent to taking into account only this first term.
The second term (inter-band contributions)
$\Delta X^2_{\NB}(E,t)$  is
a non-ballistic (non-Boltzmann) contribution.
It is
due to the non-diagonal
elements of the velocity operator and describes
the spreading of the wave-function
in a cell.
This term is
bounded by $L_{m}(E)^2$, which depends on the energy $E$ \cite{Mayou08RevueTransp}.
From numerical calculations, it is found that
$\Delta X^2_{\mathrm{NB}}(E,t)$ reaches rapidly its maximum limit $\sim L_{m}(E)^2$ 
(figure \ref{Fig_DX2_NB}).
Moreover for many energies, numerical calculations show that $L_{m}(E)$ is
of the order of magnitude of the unit cell size $L_k$ or less
(see also Ref. \cite{PRL06}). 
It suggests that in approximants the medium range order localises partially electron states 
and this phenomenon should lead to anomalous diffusion in quasicrystalline phases.
For large time $(t > L_{m}/V_{\B})$ the ballistic  contribution
dominates but for small time  $(t < L_{m}/V_{\B})$ the non-ballistic
contribution could dominate (``Small velocity regime'' \cite{PRL06}).
When the size of the approximants increases, the characteristic time 
$L_{m}/V_{\B}$ of the crossover
between ballistic and non-ballistic behaviour increases.

\begin{figure}[]
\begin{center}
\includegraphics[width=8cm]{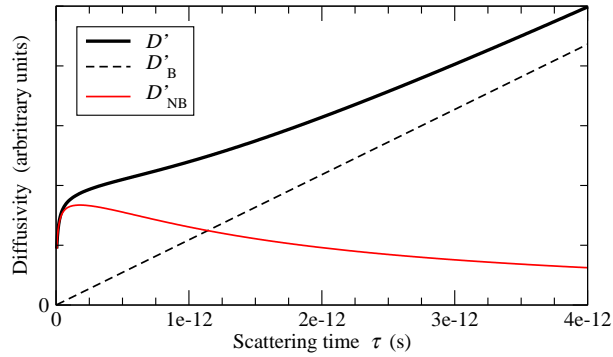}
\caption{\label{Fig_Dif_E.5}
(colour online)
Diffusivity $D'$, in relaxation time approximation, versus scattering time $\tau$,
at $E_{\rm F}=0.5$\,eV
in octagonal approximant $k=5$. $D' = D'_{\B} + D'_{\NB}$.
}
\end{center}
\end{figure}

In the relaxation time approximation \cite{Mayou00,PRL06,JubileeQC,Mayou08RevueTransp},
the role of phonons and static defects are taken into account by a scattering time $\tau$.
$\tau$ decreases when temperature $T$ increases and when the number of static defects increases.
The velocity correlation function $C'$ at temperature $T$ and/or
with static defects is computed
from $C$ at $T=0$\,K in perfect tiling,
\beq
C'(E,t) = C(E,t) \e^{-|t|/\tau}.
\eeq
Here the Fermi-Dirac distribution function is taken equal to its zero temperature value. 
This is valid provided that the electronic properties vary
smoothly on the thermal energy scale $k_{\B}T$.
The static diffusivity $D'$ is then, \cite{Mayou08RevueTransp}
\beq
D'(E_{\rm F},\tau) =\frac{1}{2}  \, \int_0^{+\infty} C(E_{\rm F},t) \e^{-t/\tau} \,{\rm d}t.
\eeq
Roughly speaking, the transport properties are almost entirely determined by the
$\Delta X^2(E,t=\tau) $.
Therefore, in large approximants for which $(\tau < L_{m}/V_{\B})$, transport properties will be
governed by non-ballistic (non-metallic) behaviour.
Figure \ref{Fig_Dif_E.5} shows the diffusivity in approximant $k=5$,
calculated in the relaxation time approximation. 
$D'$ is the sum of a ballistic term, $D'_{\B} \propto \tau$,
and a non-ballistic term, $D'_{\NB}$.
For small $\tau$ value, its behaviour is not  ballistic ($D' \neq D'_{\B}$).
In larger approximants $(k>5)$, one can expect an ``insulating like'' 
behaviour ($D'$ decreases when $\tau$ increases) for a realistic range of 
$\tau$ values, as that found in real approximants \cite{PRL06}.

\section{Conclusion}

To summarise, we have presented quantum diffusion by a pure hopping Hamiltonian in approximants 
of octagonal tiling $(k \leq 5)$.
When the size of the approximant increases the usual Boltzmann term decreases rapidly and new 
non-Boltzmann terms become essential to understand transport properties. 
These non-Boltzmann terms can have 
``insulating like'' behaviour, suggesting that in larger approximants $(k>5)$, 
``insulating like'' states,
due to long range quasiperiodic order, could exist.
Calculations in larger approximants are in progress.

\section*{Acknowledgements}
The computations have been performed at the
Service Informatique Recherche (S.I.R.),
Universit\'e de Cergy-Pontoise.
We thank Y. Costes, S.I.R.,
for computing assistance.

%
%
%
%
\bibliographystyle{tPHM}
\bibliography{octobib}






\end{document}